# Silicon formation in bulk silica through femtosecond laser engraving

*Charles M. Pépin[1], Erica Block[2], Richard Gaal[1], Julien Nillon[3], Clemens Hoenninger[3], Philippe Gillet[1], Yves Bellouard[2].*

1 - Earth & Planetary Science Lab, Ecole Polytechnique Fédérale de Lausanne (EPFL), Lausanne, Switzerland
2 - Galatea Lab, Ecole Polytechnique Fédérale de Lausanne (EPFL), Neuchâtel, Switzerland
3- Amplitude Systèmes, Bordeaux, France

**Non-linear absorption phenomena induced by controlled irradiation with a femtosecond laser beam can be used to tailor materials properties within the bulk of substrates. One of the most successful applications of this technique is the ability to fabricate three-dimensional micro-devices integrating optical, mechanical or fluid handling functions in a single substrate. In this context, amorphous $SiO_2$ is the most widely studied material.**

**Here we show that short (50-fs) femtosecond pulses induce the separation of Si and O ions in $SiO_2$ substrates, leading to the formation of micro-crystallites that we identify as pure crystalline phase of Si. Interestingly, this polymorphic phase transformation occurs in the absence of laser-induced confined microexplosion and with moderate numerical aperture. These findings not only unravel a key mechanism related to the transformation of the material and its subsequent properties, but also pave the road for the development of three-dimensional Si-rich structures embedded in a pure silica phase, eventually leading to novel disruptive approaches for fabricating three-dimensional micro-devices. For instance, one could imagine a silica-host substrate, in which arbitrary three-dimensional silicon-based components are direct-write using a femtosecond laser, rather than through assembly of components coming out of different substrates or using multiple processing steps.**

Femtosecond laser exposure of fused silica in the low-energy regime where no ablation occurs is known to produce subtle changes in the material that manifest themselves by the occurrence of densified zones (Mirua, Qui, Inouye, Mitsuyu, & Hirao, 1997) or by the appearance of self-organized porous nanoplanes forming nanogratings (Shimotsuma, Kazansky, Qui, & Hirao, 2003). So far, it was believed that these modifications were mostly related to the generation of defects such as colored centers, non-bridging oxygen hole centers or dangling bonds as well as to the formation of shortened silica rings (Chan, Huser, Risbud, & Krol, 2001) (Strelsov & Borrelli, 2002)(Shcheblanov, Povarnitsyn, Mishchik, & Tanguy, 2018).

Here, we show evidences based on Raman spectra that low-energy, short femtosecond laser pulses, not tightly focused are able to create crystalline phases within the bulk of the material. In particular, we found crystallites that can be related to a silicon phase. Noticeably these crystallites disappear when the Raman laser intensity is increased and turn back into an amorphous silica amorphous structure, suggesting that these phases occupy a limited volume, of typically a few microns and oxidize while heated up.

The existence of silicon phases formed into fused silica provides a clue for understanding a peculiar observation reported earlier by S. Matsuo *et al.* (Kyiyama, Matsuo, Hashimoto, & Morihira, 2009) where it was demonstrated that a strong etching selectivity could be achieved in a aqueous KOH solution, despite the fact that KOH normally does not etch fused silica significantly.

This remarkable observation suggests that femtosecond lasers could be used for forming tracks or predefined 3D volumes of silicon phases embedded in a glass, forming a novel type of *direct-write* composite structure with interesting optical and electrical properties. Such composite could find numerous applications, in particular for novel photonics devices combining semi-conducting phases with other structure like waveguides as well as novel electronics components.

**Methods and results**

Fused silica glass specimens of 1 mm thickness are used. A femtosecond fiber laser ($\lambda$= 800 nm, $\Delta t_{pulse}$= 50 fs, repetition rate of 120 kHz) is focused approximately 15 µm below the surface using a 0.65 NA objective and with varying linear polarization state.. The silica sample is moved in the three directions under the laser beam using computer-controlled stages. The laser-pulse energies were fixed at 580 nJ for two different runs, and the scanning speed in one direction was varied from 0.5 mm/s to 600 mm/s. This value corresponds to an instant power density of ~5 $10^{14}$ W/cm$^2$ and a net fluence of up to ~60 J/mm$^2$. Along each irradiated line the focus depth was progressively reduced using the z-stage so as to make the line progressively emerge from the surface of the plate (see Fig. 1).

After laser irradiation, the processed lines were analyzed using Raman scattering spectroscopy. The Raman spectra were collected using a LabRam Raman spectrometer equipped with a 1800 g/mm grating and a 532.19 nm laser. In the regions where the lines were emerging, micrometer-sized shining crystallites could be observed sparingly under normal illumination in the Raman (see inset of fig.2). Using very low Raman laser power, ~ 0.2mW, the collected Raman spectrum (fig. 2, red line) appears to be dramatically different than the one of bulk SiO$_2$ (fig.2, black line). Upon increasing the power of the Raman laser up to 200 mW, the spectrum of the crystallite starts to change progressively, finally transforming into the characteristic spectrum of bulk SiO$_2$ (fig.3). When collecting Raman spectra at different places along the same laser-written line - but not on a visible crystallite, a typical spectrum of femtosecond-laser irradiated silica is observed (fig.2, grey line). Noticeably, in these spectra two very strong Raman peaks at 1549 and 1556 cm$^{-1}$ are observed. These peaks can be attributed to the presence of molecular O$_2$. The peak at 1556 cm$^{-1}$ corresponds to free molecular O$_2$ (Berger, Wang, Sammeth, Itzkan, Kneipp, & Feld, 1995), while the peak at 1549 cm$^{-1}$ is attributed to dissolved O$_2$ within the silica network (Skuja & Güttler, 1996). We also observed these peaks in another work where specimens exposed to 150 fs-femtosecond pulses were analyzed (Bellouard, et al., 2016).

Stunning features can be retrieved from the new Raman spectrum (fig.2, red line). While the sharp Raman peaks are indicative of an amorphous to crystal transition, the low-frequencies at which these peaks are observed are indicative of a weak bonding typical of a metal or of a narrow-gap semiconductor. This is further supported by the shiny aspect of the crystallite, meaning that we also observe an insulator to semiconductor or metal transition. Strikingly, no molecular oxygen is observed when probing the crystallite. Added to the fact that the Si-O bond usually exhibits a strong Raman peak above 400 cm$^{-1}$, we conclude that no oxygen is present in this crystallite, implying that a chemical dissociation took place during the femtosecond laser irradiation resulting in the formation of a crystallite of pure Si.

The allotropy of silicon at ambient conditions is very rich and numerous phases have been obtained: cubic Si-I -the conventional silicon with a diamond structure and the only thermodynamically stable form at ambient conditions, metastable cubic $Si_{136}$ and orthorhombic $Si_{24}$, open-framework zeolite-like structures (Kasper, P, Pouchard, & Cros, 1965) (Kim, Stefanoski, Kurakevych, & Strobel, 2014), metastable dense cubic Si-III, with a BC8 structure (Wentorf & Kasper, 1963), hexagonal Si-IV (Wentorf & Kasper, 1963), Si-XII or R8 (Piltz, Maclean, Clark, Ackland, Hatton, & Crain, 1995), two polytypes of Si-I, called 9R and 27T (Lopez, Givan, Cornell, & Lauhon, 2011) and some new structures observed by TEM/SAED (Rapp, et al., Experimental evidence of new tetragonal polymorphs of silicon formed through ultrafast laser-induced confined microexplosion, 2015). Our Raman spectra could be explained by a mixture of Si-III and Si-XII, although the correspondence with the known Raman peaks at ambient conditions is not perfect, which may indicate the presence of residual stress which is highly probable considering the formation of these phases (i.e. femtosecond laser pulses induce strong shock-wave in the material forming high pressure conditions). We note that another candidate may be found among the numerous structures proposed by Rapp *et al.* (Rapp, et al., Experimental evidence of new tetragonal polymorphs of silicon formed through ultrafast laser-induced confined microexplosion, 2015), however, in this work, no Raman data is available for comparison. It should also be noted that in the work of Rapp *et al.*, who also used fs-laser irradiation, different focusing conditions, temporal characteristics and substrates are used. There, Si-polymorphs are generated at the interface between a silicon substrate and an oxidized layer by creating a strong shockwave resulting from a 'microexplosion' consequent to a tight focusing with a high numerical aperture objective. In their study, the polymorphism mechanism is attributed to the high-pressure shockwave generated at the interface between the $SiO_2$ layer and the pristine Si, due to plasma rapid expansion in the oxide layer. Nevertheless, it is interesting to note that all the candidate structures are synthesized at high pressure, typically around tenth of GigaPascal (1 GPa ≈ 10000 atm) and are brought back metastably at ambient conditions. This agrees with the fact that we find a permanent densification of ~4% corresponding to a maximum pressure of ~12 GPa in the irradiated silica close to the crystallites, using the method proposed by Deschamps *et al.* (Deschamps, et al., 2013). This method, taking into account a large part of the Raman spectrum and for which the determination is reproducible, offers a relationship between selective Raman parameters and the degree of densification in the glass.

Micro-crystallites were only found at places where the lines emerge at the surface i.e. where they could be optically identified. In order to look for such crystallites under the surface, the sample was cleaved perpendicularly to the writing direction where the lines are writing 6 microns below the surface (fig.4, photograph). A mild polishing was also realized during the sample preparation. Raman spectroscopy at low laser power inn the exposed section (fig. 4, red line), shows some differences when compared with a reference taken on a non-irradiated region of the same section (fig.4, green line). The new peak around 900 $cm^{-1}$ is attributed to the polishing slurry. At low-frequencies some remaining features are visible indicating that some changes indeed took place even under the surface. However, no crystallites like the ones detected at the surface were clearly observed in these sections.

**Discussion**

The formation of micrometric Si-crystallites from a pure SiO$_2$ substrate was not reported before and, to date may have been unnoticed, as it requires low power focused Raman power to prevent the rapid transformation of the crystallite through an oxidation mechanism.

Here, contrary to previous attempt to generate polymorphic phases reported before (Juodkazis, et al., 2006) (Vailionis, Gamaly, Mizeikis, Yanf, Rode, & Juodkazis, 2011)(Rapp, et al., 2015), we do not make use of confined micro-explosion but rather use moderately focused beam (numerical aperture of 0.65 versus NA of 1.45) with similar peak intensity but shorter pulses that we scan on the specimen, just like one would normally do in normal 3D printing. Using this method, we were able to synthesize micrometer-size crystalline Si out of a single SiO$_2$ specimen. These laser-exposure conditions generate well-defined modified zones, without introducing side effects, such as cavities or cracks consecutive to the use of micro-explosion as reported in other works and offer the possibility to cover or define arbitrary volume of modified materials.

All the Si allotrope candidates considered having a band gap inferior to 2.3 eV, close to the energy of the Raman laser, the crystallite strongly absorbs the laser and starts to heat up as the power increases. Looking at the Raman sequence, it most likely induces an oxidation reaction in contact with the ambient atmosphere and transforms back to amorphous SiO$_2$. Only a careful investigation of the irradiated products at very low Raman laser power allowed us to notice these micrometric crystals. Finding a silicon phase consecutive to the laser exposure is consistent with the formation of molecular oxygen in the laser-affected zones (as shown experimentally in (Bellouard, et al., 2016) and proposed speculatively in Lancry and Poumellec (Lancry, Poumellec, Canning, Cook, Poulin, & Brisset, 2013)). As the silica matrix is depleted from oxygen atoms, Si-Si bonds formation is likely, eventually forming Si polymorphs. Finally, the formation of these Si phase offers a plausible explanation of two puzzling, yet unexplained observations, reported earlier. The first one is related to the enhanced etching by KOH of laser-affected zones as reported by Kiyama *et al. (Kyiyama, Matsuo, Hashimoto, & Morihira, 2009)*. In normal situation, KOH does not affect silica and the strong etching enhancement reported by Kiyama *et al.* cannot be explained by densification of the silica matrix. The presence of a silicon phases offers a consistent explanation to this phenomenon. The second puzzling fact was the recent report of weak surface enhancement mechanism (SERS) (Bellouard, Block, Squier, & Gobet, 2017) observed in ejecta from silica ablated lines under similar sub-100fs femtosecond laser exposure (using low NA ~0.05, simultaneous spatial temporal focusing of 85 fs pulses at 1 kHz, up to 80 uJ). As these nanoparticles are likely to be enriched with Silicon nanocrystals, similar to the ones we are detecting here, with the consequence of exhibiting conducting or semi-conducting properties considering the small bandgap, it may account for the SERS enhancement mechanism observed.

The discovery of a silicon crystallization induced by femtosecond laser irradiation is of great interest for the SiO$_2$ processing. Silicon being the archetype of a semiconductor and the basic constituent of a wealth of photonics and electronics components, the prospect of direct-writing three-dimensional silicon components with arbitrary shapes, self-packaged in a high purity silica matrix opens up an entirely new field of investigations for highly efficient optoelectronics devices fully exploiting the three-dimensions. Considering the ubiquity of silicon-based devices in our daily life, disruptive applications are foreseeable, in particular considering the prospects offer by three-d printing methods and the possibility of seamless integration of multiple functionalities combining mechanics, fluid-handling and optical functions in a same substrate.

Although further follow-up studies are needed to further characterize theses laser-induced silicon nanostructures, this work defines a milestone as it clearly demonstrates that these lasers are able to induce phase separation at a nanoscale, turning locally silica into silicon.

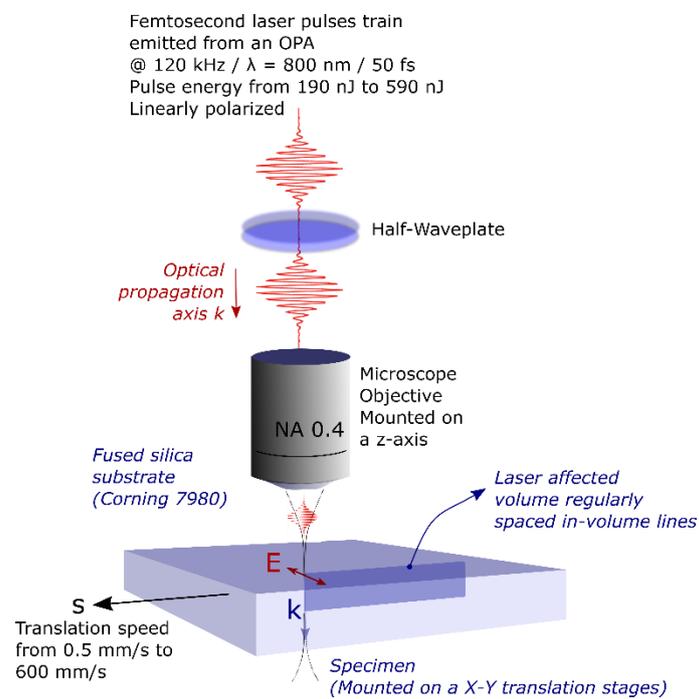

**Figure 1 – Schematic of the experimental setup. A glass specimen mounted on translation stages is moved under a laser beam focused within the specimen volume. The writing speed, the pulse energy and polarization state of the beam are tuned to vary the material exposure dose as well as the rate of energy deposition.**

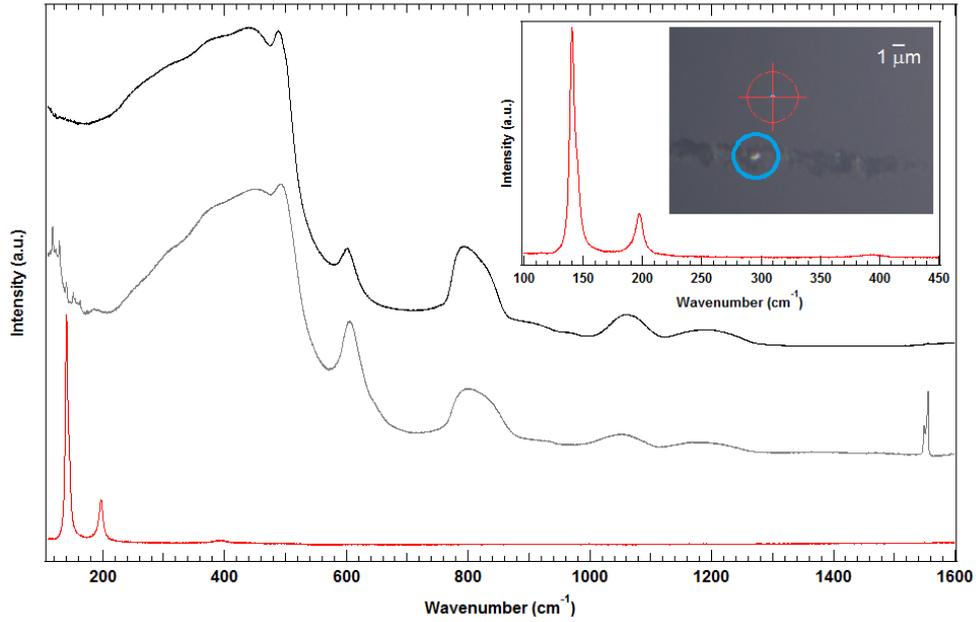

**Figure 2 -** Raman spectra of the new femtosecond laser-induced phase (red line) compared with a spectrum taken at another location along the same line (grey line) and with a reference taken in bulk SiO2 (black line) Inset: zoom of the low-frequency part of the Raman spectrum of the new phase, taken on the micro-crystallite highlighted by the blue circle on the picture

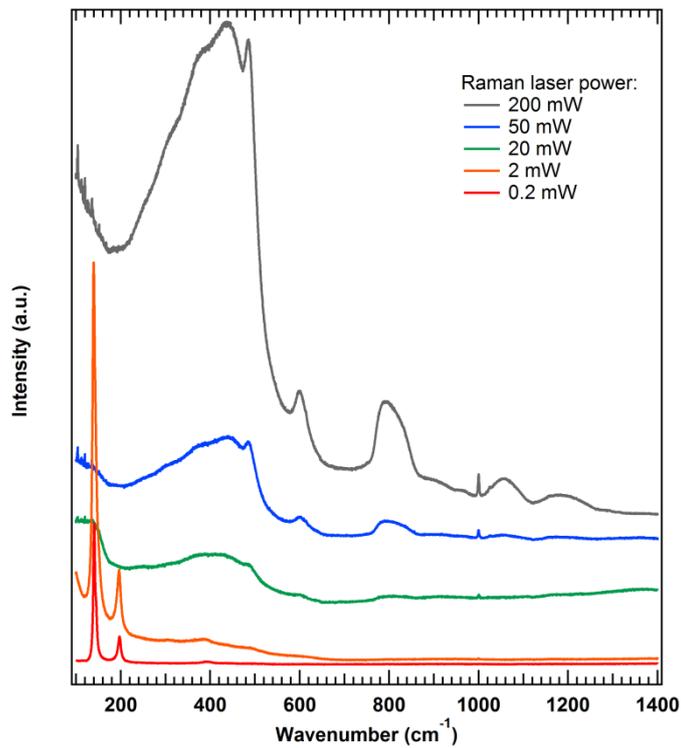

**Figure 3 -** Progressive change in the Raman spectrum of the new phase as the intensity of the Raman laser is increased. The sharp peaks at low wavenumbers observed for very low Raman laser power are indicative of a new crystalline phase, which progressively transforms back to amorphous $SiO_2$.

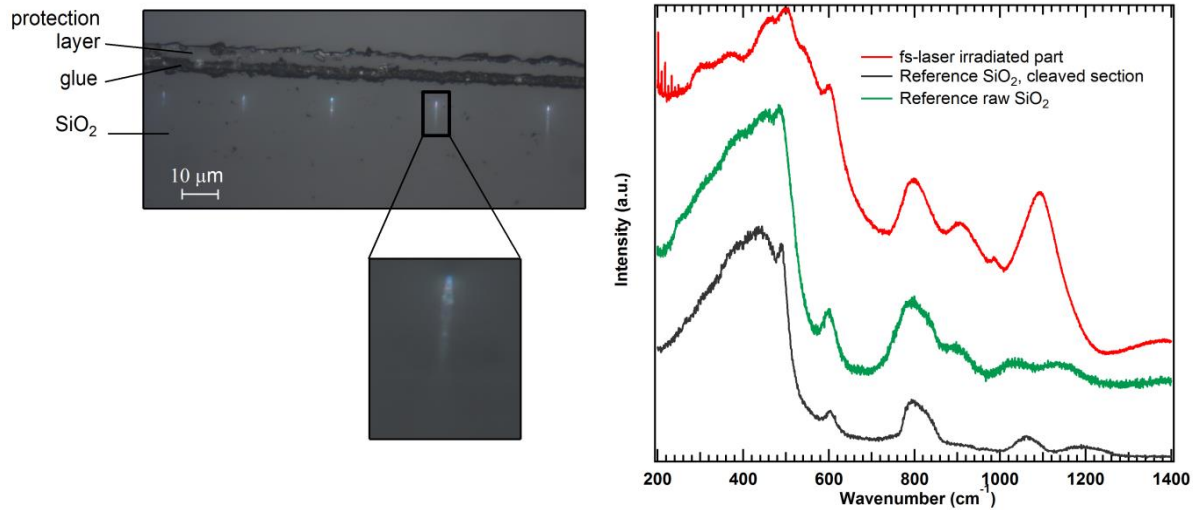

Figure 4 - Left: Photograph of the cleaved sample. Lines written 6 microns below the surface are visible. Before cleaving it, the sample is protected by a small layer glued to the surface and still visible on the photograph. Right: Raman spectra taken on the freshly cleaved section of the sample. Black line: reference in bulk silica, green line: reference on the cleaved section, red line: spectrum taken on the femtosecond laser irradiated part